\documentclass[aps, prd, superscriptaddress, eqsecnum, nofootinbib, showkeys, twocolumn, showpacs, preprintnumbers, amsmath, amssymb]{revtex4-1}

\usepackage[usenames, dvipsnames]{xcolor}

\usepackage{graphicx}

\usepackage{gensymb}

\usepackage{tabularx}

\begin{document}

\title{Suppression of long-wavelength CMB spectrum from the no-boundary initial condition}

	\author{Pisin Chen}
		\email{pisinchen@phys.ntu.edu.tw}
		\affiliation{Department of Physics, National Taiwan University, Taipei 10617, Taiwan}
		\affiliation{Leung Center for Cosmology and Particle Astrophysics, National Taiwan University, Taipei 10617, Taiwan}
		\affiliation{Kavli Institute for Particle Astrophysics and Cosmology, SLAC National Accelerator Laboratory, Stanford University, Stanford, California 94305, USA}
		\affiliation{Graduate Institute of Astrophysics, National Taiwan University, Taipei 10617, Taiwan}
	
	\author{Yu-Hsiang Lin}
		\email{d00222001@ntu.edu.tw}
		\affiliation{Department of Physics, National Taiwan University, Taipei 10617, Taiwan}
		\affiliation{Leung Center for Cosmology and Particle Astrophysics, National Taiwan University, Taipei 10617, Taiwan}
	
	\author{Dong-han Yeom}
		\email{innocent.yeom@gmail.com}
		\affiliation{Leung Center for Cosmology and Particle Astrophysics, National Taiwan University, Taipei 10617, Taiwan}
     	\affiliation{The Asia Pacific Center for Theoretical Physics, Pohang 37673, Korea}
	     \affiliation{Department of Physics, POSTECH, Pohang 37673, Korea}
	\date{\today}


\begin{abstract}

	The lack of correlations at the long-wavelength scales of the cosmic microwave background spectrum is a long-standing puzzle and it persists in the latest Planck data. By considering the Hartle-Hawking no-boundary wave function as the initial condition of the inflationary universe, we propose that the power suppression can be the consequence of a massive inflaton, whose initial vacuum is the Euclidean instanton in a compact manifold. We calculate the primordial power spectrum of the perturbations, and find that as long as the scalar field is moderately massive, the power spectrum is suppressed at the long-wavelength scales.

\end{abstract}

\maketitle


\section{Introduction}
\label{sec:Introduction}
	
	Thanks to numerous cosmological observations, now we can investigate the universe with high precisions. One of the most important observational objects for the precision cosmology is the cosmic microwave background (CMB). The recent observational result of the CMB two-point function from the Planck mission \cite{Planck2016XX} is well described by the $\Lambda$CDM model and the single-field inflation scenario (to which we refer as the ``standard scenario'' hereafter). This demonstrates a great success of the inflation scenario \cite{Guth1981,Linde1983,Starobinsky1980}.
	
	However, it is also fair to say that the observed two-point correlation function at long-wavelength scales has a statistical tension with the standard scenario. More precisely, the observed spectrum at the long-wavelength scales shows a lack of correlation \cite{Planck2016XI}. Although such a tension is not statistically significant yet, it can be confirmed or falsified by upcoming experiments. In either conclusion, the power suppression problem can shed lights on the physics beyond the inflationary cosmology.
	
	If the power suppression at long-wavelength scales is confirmed by future observations, what will be the cause of it? One candidate is the physics at the beginning of the inflation. Within the context of the semi-classical quantum field theory and general relativity, the power suppression can occur if one of the two following possibilities happens in the early stage of the inflation \cite{Chen2016}. First, the phantom equation of state (and the super-inflationary expansion due to the phantomness) can induce the power suppression. Second, a positive-pressure era (with the equation-of-state parameter $w > 0$), such as the kinetic-energy-dominated era, at the early stage of inflation can cause the power suppression. Both scenarios are logically possible, but both ideas have their own problems. For the phantom inflation scenario, it is very difficult to construct a viable theory for the phantom matter. For the positive-pressure era, the power suppression highly depends on the choice of the vacuum state. In the de Sitter space, we have a canonical choice of the vacuum---the Bunch-Davies vacuum \cite{Bunch1978}, but in the positive-pressure era, there is no such a canonical vacuum. Moreover, if we consider an eternally inflating background (and the consequent Bunch-Davies vacuum), then even though the universe evolves toward a positive-pressure era, the power suppression will not be realized \cite{Chen2016}.
	
	The existing difficulties of having a consistent explanation for the power suppression may imply that its origin does not lie in the semi-classical physics, but in the quantum theory of gravity. \emph{Can we explain the power suppression by quantum gravitational effects?} Indeed, there has been several models explaining the power suppression from quantum gravity \cite{Kiefer:2011cc,Bini:2013fea,Kamenshchik:2014kpa,Kamenshchik:2015gua,Brizuela:2015tzl,Kamenshchik:2016mwj,Brizuela:2016gnz,Kamenshchik:2017kfs,Kamenshchik:2018rpw} although almost all discussions have been limited to the flat Friedmann-Robertson-Walker universe. For example, according to the loop quantum cosmology, quantum gravitational effects can induce an effective phantom matter in the deep trans-Planckian regime. The phantomness thereof can explain the CMB power suppression as well as supporting the scenario of the big bounce universe \cite{Ashtekar2015}.
	
	In order to investigate the wave function of our universe and the power suppression problem, we will rely on the Hartle-Hawking wave function, or the so-called no-boundary wave function \cite{Hartle1983}. This wave function is one of the proposals to the boundary condition of the Wheeler-DeWitt equation \cite{DeWitt1967}. It is a path integral over the Euclidean compact manifolds, and can be approximated by the method of steepest descent. Under such approximation, we can then describe the wave function as a sum of the Euclidean instantons, where each instanton should eventually be Wick-rotated into the Lorentzian signatures \cite{Hartle2008a, Hartle2008} and approach real-valued functions \cite{Hwang2012, Hwang2012a, Hwang2014, Hwang2015, Chen2016a}. By integrating the Lagrangian, one can estimate the probability for the history described by each instanton.
	
	Following the work of Halliwell and Hawking \cite{Halliwell1985}, one can introduce perturbations to the background instanton solution. These perturbations also carry their own canonical degrees of freedom. Although in general it is very difficult to track their coupled evolution, one can consistently consider various modes separately as long as the perturbations stay in the linear regime. The probability distribution of the magnitude of each perturbation mode can then be calculated, and the expectation values of these modes, or equivalently, the power spectrum, can therefore be determined.
	
	In this paper, we devote several sections to revisit the formalism of Halliwell and Hawking. Using the method of Laflamme \cite{Laflamme1987}, we can define the wave function for the Euclidean vacuum. The Euclidean vacuum gives the scale-invariant power spectrum at short-wavelength scales, hence consistent with the choice of the Bunch-Davies vacuum \cite{Bunch1978} at small scales. On the other hand, at the long-wavelength scales, the power spectrum is enhanced due to the curvature of the manifold. All these results have been known in the literature and consistent with the independent calculations from quantum field theoretical techniques \cite{Starobinsky1996, Halliwell1990}. However, to our best knowledge, it was not emphasized that the power spectrum can be \textit{suppressed} by introducing the potential term. In this paper, we include analytical and numerical details for the power suppression due to the potential term of the inflaton field.
	
	The paper is organized as follows. We introduce the minisuperspace model and the no-boundary wave function in Sec.~\ref{sec:Background}. We calculate the contributions from the perturbations and the power spectrum in Sec.~\ref{sec:Perturbation}. We solve the equations of motion of the perturbations and investigate the effect of the mass of the scalar field in Sec.~\ref{sec:Solution}. We conclude in Sec.~\ref{sec:Conclusion}.
	
	We use the Planck units ($\hbar = c = G = 1$) in this paper.



\section{Minisuperspace model}
\label{sec:Background}
	
	In this section, we describe the Hartle-Hawking wave function in the minisuperspace model \cite{Hartle2008a, Hartle2008}. Especially, we focus on the background-level solution.
	
	The ADM metric for the homogeneous closed universe is
		\begin{align}
			ds^2 &= \sigma^2 \left[ -(\bar{N}^2 - \bar{N}_i \bar{N}^i) d\lambda^2 + 2 \bar{N}_i dx^i d\lambda \right. \notag \\
			&\quad \left. + \bar{h}_{i j} dx^i dx^j \right],
		\end{align}
		where $\sigma$ is a constant normalization, and
		\begin{align}
			\bar{N} &= N_0(\lambda), \\
			\bar{N}_i &= 0, \\
			\bar{h}_{i j} &= a^2(\lambda) \bar{\gamma}_{i j}, \\
			\bar{\gamma}_{i j} dx^i dx^j &= d\chi^2 + \sin^2\chi (d\theta^2 + \sin^2\theta d\varphi^2) = d\Omega_3^2.
		\end{align}
		The action for a scalar field in the close universe is
		\begin{align}
			I &= \frac{1}{16 \pi} \int d^4x \sqrt{-g} R \notag \\
			&\quad + \int d^4x \sqrt{-g} \left[ -\frac{1}{2} \partial^{\mu} \Phi \partial_{\mu} \Phi - V(\Phi) \right],
		\end{align}
		where
		\begin{align}
			%
			V(\Phi) = V_0 + \frac{1}{2} m^2 \Phi^2.
		\end{align}
		Defining the variables
		\begin{align}
			\phi &= \sqrt{\frac{4 \pi}{3}} \; \Phi, \\
			\tilde{V}(\Phi) &= \frac{8 \pi \sigma^2}{3} V(\Phi),
		\end{align}
		and integrating over the compact geometry, the action can be expanded as
		\begin{align}
			I[N_0, a, \phi] &= \frac{3 \pi \sigma^2}{4} \int d\lambda \; N_0 \left\{ -a \left( \frac{a'}{N_0} \right)^2 + a \right. \notag \\
			&\quad \left. + a^3 \left[ \left( \frac{\phi'}{N_0} \right)^2 - \tilde{V}(\Phi) \right] \right\},
		\end{align}
		where the primes denote the derivatives against $\lambda$. It is convenient to further define
		\begin{align}
			\tilde{V}_0 &= \frac{8 \pi \sigma^2}{3} V_0, \\
			\tilde{m} &= \sigma m,
			%
			%
		\end{align}
		so that
		\begin{align}
			%
			\tilde{V} = \tilde{V}_0 + \tilde{m}^2 \phi^2.
		\end{align}
	
	The no-boundary wave function can be written as the path integral,
		\begin{align}
			\label{eq:WaveFunction} \Psi = \int \mathcal{D} \hat{a} \mathcal{D} \hat{\phi} \mathcal{D} \hat{N} \; e^{-\frac{1}{\hbar} \hat{I}[ \hat{a}, \hat{\phi}, \hat{N} ]},
		\end{align}
		where $\hat{a}$, $\hat{\phi}$, and $\hat{N}$ are the corresponding fields in the Euclidean metric,
		\begin{align}
			ds^2 = \sigma^2 \left[ \hat{N}_0^2 d\lambda^2 + \hat{a}^2 d\Omega_3^2 \right],
		\end{align}
		obtained from the Lorentzian one by substituting $N_0$ by $-i \hat{N}_0$ and adding hats to other fields for clarity. The Euclidean action is taken as $\hat{I} = -i I|_{N_0 = -i \hat{N}_0}$:
		\begin{align}
			\hat{I} &= \frac{1}{16 \pi} \int d^4 x \sqrt{+\hat{g}} \hat{R} \notag \\
			&\quad + \int d^4 x \sqrt{+\hat{g}} \left[ -\frac{1}{2} \partial^{\mu} \hat{\Phi} \partial_{\mu} \hat{\Phi} - V(\hat{\Phi}) \right].
		\end{align}
		After integration, we have
		\begin{align}
			\hat{I}[\hat{N}_0, \hat{a}, \hat{\phi}] &= \frac{3 \pi \sigma^2}{4} \int d\lambda \; \hat{N}_0 \left\{ -\hat{a} \left( \frac{\hat{a}'}{\hat{N}_0} \right)^2 - \hat{a} \right. \notag \\
			&\quad \left. + \hat{a}^3 \left[ \left( \frac{\hat{\phi}'}{\hat{N}_0} \right)^2 + \tilde{V}(\hat{\Phi}) \right] \right\}.
		\end{align}
	
	Doing variation with respect to $\hat{N}_0$, we obtain the Hamiltonian constraint,
		\begin{align}
			%
			\label{eq:EuclideanHamiltonianConstraint} \dot{\hat{a}}^2 - 1 + \hat{a}^2 \left[ - \dot{\hat{\phi}}^2 + \tilde{V}(\hat{\Phi}) \right] = 0,
		\end{align}
		where dots denote derivatives against $\tau$, which is defined by
		\begin{align}
			d\tau = \hat{N}_0 d\lambda.
		\end{align}
		Using the steepest descent approximation, the wave function is dominated by the extreme path $(\hat{a}_{\textrm{ext}}(\tau), \hat{\phi}_{\textrm{ext}}(\tau))$ that satisfies
		\begin{align}
			\frac{\delta \hat{I}}{\delta \hat{a}} &= 0, \\
			\frac{\delta \hat{I}}{\delta \hat{\phi}} &= 0,
		\end{align}
		which are
		\begin{align}
			\label{eq:EuExt1} \ddot{\hat{a}} + 2 \hat{a} \dot{\hat{\phi}}^2 + \hat{a} \tilde{V}(\hat{\Phi}) &= 0, \\
			\label{eq:EuExt2} \ddot{\hat{\phi}} + 3 \frac{\dot{\hat{a}}}{\hat{a}} \dot{\hat{\phi}} - \frac{1}{2} \frac{\partial \tilde{V}}{\partial \hat{\phi}} &= 0,
		\end{align}
		respectively. Note that the Hamiltonian constraint \eqref{eq:EuclideanHamiltonianConstraint} is used when deriving the equations above.
	
	To solve $\hat{a}(\tau)$, we consider the case in which $\dot{\hat{\phi}}^2$ is negligible, and combine \eqref{eq:EuclideanHamiltonianConstraint} and \eqref{eq:EuExt1} to obtain
		\begin{align}
			\label{eq:aEqSlowRoll} \hat{a} \ddot{\hat{a}} - \dot{\hat{a}}^2 + 1 = 0.
		\end{align}
		The ``no-boundary'' boundary condition at $\tau = 0$ sets $\hat{a}(0) = 0$. To keep \eqref{eq:EuExt2} finite, we also require $\dot{\hat{\phi}}(0) = 0$. Then by the Hamiltonian constraint \eqref{eq:EuclideanHamiltonianConstraint} we know $\dot{\hat{a}}(0) = 1$. The only free initial conditions left are the real and imaginary parts of $\hat{\phi}(0)$.
	
	Equation \eqref{eq:aEqSlowRoll} has four solutions,
		\begin{align}
			\hat{a}(\tau) &= \pm \frac{1}{H_0} \sin\left[ H_0 ( \tau - \tau_0 ) \right], \\
			\hat{a}(\tau) &= \pm \frac{1}{H_0} \sinh\left[ H_0 ( \tau - \tau_0 ) \right].
		\end{align}
		For physical solutions we should pick the plus sign. By requiring $\hat{a}(0) = 0$ we have $\tau_0 = 0$, and automatically we have consistently $\dot{\hat{a}}(0) = 1$. In order to connect to the Lorentzian space, which requires $\hat{a}'(\tau_{\textrm{connect}}) = 0$, the qualified solution is
		\begin{align}
			\label{eq:scaleAEu} \hat{a}(\tau) = \frac{1}{H_0} \sin( H_0 \tau ).
		\end{align}
	
	The solution connects to the Lorentzian space at $\tau_{\textrm{connect}} = \pi / 2 H_0$. In Lorentzian space, we define
		\begin{align}
			dt = N_0 d\lambda,
		\end{align}
		therefore $d\tau = i dt$. We can then describe the Euclidean trajectory by $\tau = 0$ to $\pi / 2 H_0$, and the Lorentzian one by the complex contour
		\begin{align}
			\tau = \frac{\pi}{2 H_0} + i t
		\end{align}
		with $t > 0$. We then have the Lorentzian solution
		\begin{align}
			\label{eq:scaleALo} a(t) = \frac{1}{H_0} \cosh( H_0 t ).
		\end{align}
		%


\section{Perturbation spectrum from the wave function}
\label{sec:Perturbation}

In this section, we include the perturbations of the matter field as well as the metric on top of the background-level solution. By using the steepest decent approximation again, we can calculate the expectation values of perturbations. This section is a revisit of the paper of Halliwell and Hawking \cite{Halliwell1985}.
	
	The perturbations to the spatial part of the metric in the $S^3 \times R$ closed universe can be organized as
		\begin{align}
			h_{i j} &= a^2 \gamma_{i j}, \notag \\
			\gamma_{i j} &= \bar{\gamma}_{i j} + \epsilon_{i j},
		\end{align}
		where $\epsilon_{i j}$ denotes
		\begin{align}
			\label{eq:epsilon} \epsilon_{i j} &= \sum_{n, l, m} \left[ \sqrt{6} q_{n l m} \frac{1}{3} \bar{\gamma}_{i j} Q_{n l m} + \sqrt{6} b_{n l m} ( P_{i j} )_{n l m} \right. \notag \\
			&\quad\quad\quad\quad + \sqrt{2} c^o_{n l m} ( S^o_{i j} )_{n l m} + \sqrt{2} c^e_{n l m} ( S^e_{i j} )_{n l m} \notag \\
			&\quad\quad\quad\quad \left. + 2 d^o_{n l m} ( G^o_{i j} )_{n l m} + 2 d^e_{n l m} ( G^e_{i j} )_{n l m} \right],
		\end{align}
		and
		\begin{align}
			P_{i j} = \frac{1}{n^2 - 1} \nabla_i \nabla_j Q + \frac{1}{3} \bar{\gamma}_{i j} Q.
		\end{align}
		Here the covariant derivatives are with respect to $\bar{\gamma}_{i j}$. The first, second, and third lines of \eqref{eq:epsilon} denote the scalar, vector, and tensor perturbations, respectively. Suppressing the spherical coordinate indices, $n$, $l$, $m$, the coefficients, $q$, $b$, $c^o$, $c^e$, $d^o$, $d^e$, are time dependent, while the basis, $Q$, $P_{i j}$, $S^o_{i j}$, $S^e_{i j}$, $G^o_{i j}$, $G^e_{i j}$, are space dependent.
	
	The perturbations to the lapse and the shift functions are
		\begin{align}
			N &= N_0 \left[ 1 + \frac{1}{\sqrt{6}} g_{n l m} Q_{n l m} \right], \\
			N_i &= a \left[ \frac{1}{\sqrt{6}} k_{n l m} (P_i)_{n l m} + \sqrt{2} j_{n l m} (S_i)_{n l m} \right],
		\end{align}
		where
		\begin{align}
			P_i = \frac{1}{n^2 - 1} \nabla_i Q.
		\end{align}
		Finally, the perturbation to the scalar field is 
		\begin{align}
			\Phi = \sqrt{\frac{3}{4 \pi}} \phi + \sqrt{\frac{3 \pi}{2}} f_{n l m} Q_{n l m}.
		\end{align}
		Among the perturbations, $g_{n l m}$, $k_{n l m}$, and $f_{n l m}$ are the scalar ones, while $j_{n l m}$ is the vector one.
	
	The action can be expanded around the background fields to the second order as the sum of the eigenmodes \cite{Halliwell1985},
		\begin{align}
			I &= I_0(a, \bar{\phi}, N_0) \notag \\
			&\quad + \sum_{n, l, m} I_{n l m}(a, \bar{\phi}, N_0; q_{n l m}, \dots, k_{n l m}).
		\end{align}
		Choosing the gauge in which $q_{n l m} = b_{n l m} = 0$, the constraint equations can be obtained by variating the quadratic part of the perturbation action with respect to $g_{n l m}$ and $k_{n l m}$,
		\begin{align}
			\label{eq:gnlm} g_{n l m} &= 3 \frac{(n^2-1) H \dot{\phi} f_{n l m} + \dot{\phi} \dot{f}_{n l m} + \tilde{m}^2 \phi f_{n l m}}{(n^2 - 4) H^2 + 3 \dot{\phi}^2}, \\
			k_{n l m} &= 3 ( n^2 - 1 ) N_0 a \notag \\
			&\hspace{-1cm} \times \frac{H \dot{\phi} \dot{f}_{n l m} + H \tilde{m}^2 \phi f_{n l m} - 3 \dot{\phi} ( -H^2 + \dot{\phi}^2 ) f_{n l m}}{(n^2 - 4) H^2 + 3 \dot{\phi}^2}.
		\end{align}
		Here the dots denote derivatives against the Lorentzian time $t$. The equation of motion for $f_{n l m}$ can be obtained by the variation with respect to $f_{n l m}$,
		\begin{align}
			\label{eq:fEOM} &\quad\; \ddot{f}_{n l m} + 3 H \dot{f}_{n l m} + \left( \tilde{m}^2 + \frac{n^2 - 1}{a^2} \right) f_{n l m} \notag \\
			&= -2 \tilde{m}^2 \phi g_{n l m} + \dot{\phi} \dot{g}_{n l m} - \frac{\dot{\phi} k_{n l m}}{N_0 a}.
		\end{align}
	
	The amplitude of the perturbations of the scalar field, $\delta \Phi$, can be obtained through calculating the expectation value with the no-boundary wave function, focusing on the part relevant for $f_{n l m}$. Using the steepest descent approximation, the Euclidean action $\hat{I}$ in the wave function receives contributions mostly from the solution to the equations of motion, evaluated to be
		\begin{align}
			\hat{I} &\approx \frac{a^3}{2 i N} \left( f_{n l m} \frac{d f_{n l m}}{d \tau} - \frac{d \phi}{d \tau} g_{n l m} f_{n l m} \right) \notag \\
			&= \frac{1}{2} a^3 \left( f_{n l m} \frac{d f_{n l m}}{d t} - \frac{d \phi}{d t} g_{n l m} f_{n l m} \right).
		\end{align}
		We therefore have
		\begin{align}
			\Psi[f_{n l m}] \approx B_{n l m} \exp\left[ -\frac{1}{2} a^3 \left( f_{n l m} \dot{f}_{n l m} - \dot{\phi} g_{n l m} f_{n l m} \right) \right],
		\end{align}
		where the dots denote derivatives against $t$. The normalization can be fixed by requiring
		\begin{align}
			&|B_{n l m}|^2 \int_{-\infty}^{\infty} df_{n l m} \notag \\
			&\quad \left| \exp\left[ -\frac{1}{2} a^3 \left( f_{n l m} \dot{f}_{n l m} - \dot{\phi} g_{n l m} f_{n l m} \right) \right] \right|^2  = 1.
		\end{align}
		The expectation of the field perturbations averaged over the space is given by
		\begin{align}
			\langle \delta \Phi^2(t, \vec{x}) \rangle &= \frac{1}{2 \pi^2} \int d\chi d\theta d\varphi \; \sin^2\chi \sin\theta \notag \\
			&\quad\;\; \times \frac{3 \pi}{2} \sum_{n l m} \sum_{n' l' m'} \langle f_{n l m} f_{n' l' m'} \rangle Q_{n l m}  Q_{n' l' m'} \notag \\
			&= \frac{3}{4 \pi} \sum_{n l m} \langle f_{n l m}^2 \rangle,
		\end{align}
		where
		\begin{align}
			\label{eq:f2Integral} &\langle f_{n l m}^2 \rangle = |B_{n l m}|^2 \int_{-\infty}^{\infty} df_{n l m} \notag \\
			&\quad f_{n l m}^2 \left| \exp\left[ -\frac{1}{2} e^{3 \alpha} \left( f_{n l m} \dot{f}_{n l m} - \dot{\phi} g_{n l m} f_{n l m} \right) \right] \right|^2.
		\end{align}
		Defining the power spectrum, $P(n)$, as
		\begin{align}
			\langle \delta \Phi^2(t, \vec{x}) \rangle &= \sum_n \frac{n}{n^2 - 1} P(n),
		\end{align}
		with $l$ and $m$ summed over, we then find
		\begin{align}
			\label{eq:PowerSpectrum} P(n) = \frac{3 (n^2 - 1)}{4 \pi n} \sum_{l, m} \langle f_{n l m}^2 \rangle.
		\end{align}
		Note that if $\langle f_{n l m}^2 \rangle$ depends only on $n$, the summation over $l$ and $m$ can be immediately carried out, leaving (assuming $n \gg 1$)
		\begin{align}
			P(n) \simeq \frac{3 n^3}{4 \pi} \langle f_{n}^2 \rangle.
		\end{align}
	
	To evaluate the expectation value $\langle f_{n l m}^2 \rangle$, we adopt the proposal of \cite{Laflamme1987}. We replace $\dot{f}_{n l m}$ by a combination of the canonical variable $f_{n l m}$ and its c-number value $\tilde{f}_{n l m}$,
		\begin{align}
			\dot{f}_{n l m} \rightarrow \frac{\dot{\tilde{f}}_{n l m}}{\tilde{f}_{n l m}} f_{n l m}.
		\end{align}
		When $\dot{\phi}$ or the metric perturbations $g_{n l m}$ are negligible, the wave function is then
		\begin{align}
			\Psi[ f_{n l m} ] = B_{n l m} \exp\left( -\frac{a^3 \dot{\tilde{f}}_{n l m}}{2 \tilde{f}_{n l m}} f_{n l m}^2 \right).
		\end{align}
		The normalization is evaluated as
		\begin{align}
			| B_{n l m} |^2 = \sqrt{\frac{a^3 \dot{\tilde{f}}_{n l m}}{\pi \tilde{f}_{n l m}}}.
		\end{align}
		The expectation value can then be found to be
		\begin{align}
			\label{eq:f2Exp} &\langle f_{n l m}^2 \rangle = \frac{\tilde{f}_{n l m}}{2 a^3 \dot{\tilde{f}}_{n l m}}.
		\end{align}


\section{Effect of mass on the power spectrum}
\label{sec:Solution}

	In the Euclidean space, we consider the scale factor solution \eqref{eq:scaleAEu} and a constant scalar field in the background. Neglecting the metric perturbations $g_n$ (we suppress the indices $l$ and $m$ in this section, since the equation of motion does not depend on them), we calculate the field perturbations $f_n$ (we ignore the tilde that denotes the c-number solution wherever no confusion arises) by numerically solving the equation of motion
		\begin{align}
			\label{eq:fEOMNoMetricEu} \frac{d^2 \hat{f}_n}{d \tau^2} + 3 H_0 \cot(H_0 \tau) \frac{d \hat{f}_n}{d \tau} - \left[ \tilde{m}^2 + \frac{(n^2 - 1) H_0^2}{\sin^2 (H_0 \tau)} \right] \hat{f}_n = 0,
		\end{align}
		where we use the hat to emphasize that it is the solution in the Euclidean space. In order to keep equation \eqref{eq:fEOMNoMetricEu} finite, we require that both $\hat{f}_n(\tau)$ and $\hat{f}_n'(\tau)$ vanish at $\tau = 0$. More precisely, we adopt the following ansatz as the initial condition for numerical calculations:
		\begin{align}
			\hat{f}_n(\tau_i) &= \frac{1}{2} \epsilon \tau_i^2, \\
			\hat{f}_n'(\tau_i) &= \epsilon \tau_i,
		\end{align}
		where $\tau_i \ll 1$ is the initial Euclidean time from which we start to integrate the differential equations, and $\epsilon$ is an arbitrary parameter. Note that since the expectation value $\langle f_n^2 \rangle$ depends only on the ratio $\tilde{f}_n / \dot{\tilde{f}}_n$, the power spectrum is independent of the choice of $\epsilon$. In our numerical calculation, we set $\tau_i = 10^{-4}$ and $\epsilon = 1$, and evolve the Euclidean system from $\tau_i$ to $\tau_f = \pi / 2 H_0$.
	
	In the Lorentzian spacetime, we use the analytical solution \eqref{eq:scaleALo} for the scale factor and a constant scalar field in the background to model the slow-roll inflation. The equation of motion for the field perturbation reads
		\begin{align}
			\label{eq:fEOMNoMetricLo} \frac{d^2 f_n}{d t^2} + 3 H_0 \tanh(H_0 t) \frac{d f_n}{d t} + \left[ \tilde{m}^2 + \frac{(n^2 - 1) H_0^2}{\cosh^2 (H_0 t)} \right] f_n = 0.
		\end{align}
		The boundary conditions connecting the Euclidean and Lorentzian solutions are \cite{Hwang2012, Hwang2012a, Hwang2014, Hwang2015, Chen2016a}
		\begin{align}
			\textrm{Re}\{ f(t_i) \} &= \textrm{Re}\{ \hat{f}(\tau_f) \}, \\
			\textrm{Im}\{ f(t_i) \} &= \textrm{Im}\{ \hat{f}(\tau_f) \}, \\
			\textrm{Re}\{ f'(t_i) \} &= -\textrm{Im}\{ \hat{f}'(\tau_f) \}, \\
			\textrm{Im}\{ f'(t_i) \} &= \textrm{Re}\{ \hat{f}'(\tau_f) \},
		\end{align}
		where we set $t_i = 0$ to be the initial time of integration in the Lorentzian space. We then solve the system from $t_i$ to the horizon-exit time,
		\begin{align}
			t_{\textrm{exit}} = \frac{1}{H_0} \sinh^{-1} n,
		\end{align}
		for mode $n$. Note that for each mode, the expectation value \eqref{eq:f2Exp}, hence the power spectrum \eqref{eq:PowerSpectrum}, is evaluated at its horizon-exit time.


	The Hubble parameter in the Lorentzian space is
		\begin{align}
			H(t) = H_0 \tanh(H_0 t).
		\end{align}
		Therefore $H_0$ corresponds to the Hubble constant during the exponentially growing period. To fix the value of $H_0$, we consider the Hamiltonian constraint in Lorentzian space,
		\begin{align}
			\frac{\dot{a}^2}{a^2} = H^2 = \frac{8 \pi \sigma^2}{3} V(\Phi) - \frac{1}{a^2} + \frac{4 \pi}{3} \dot{\Phi}^2.
		\end{align}
		For the case that the scalar field is massless, the constant potential, $V(\Phi) = V_0$, drives the exponential growth of the scale factor $a$. The Hubble parameter is approximately
		\begin{align}
			H \approx \sqrt{ \frac{8 \pi \sigma^2}{3} V_0 }.
		\end{align}
		We choose the normalization of the metric to be
		\begin{align}
			\label{eq:sigmaConvention} \sigma^2 = \frac{1}{V_0}.
		\end{align}
		Therefore, during the exponential growth, $H \approx H_0 \approx \sqrt{8 \pi / 3}$.


	For the case of massive scalar field, during the exponential expanding period, the Hubble parameter is approximately
		\begin{align}
			H \approx \sqrt{ \frac{8 \pi}{3} \left( 1 + \frac{m^2 \Phi^2}{2 V_0} \right) }.
		\end{align}
		Note that $\tilde{m} = m / \sqrt{V_0}$ with the choice of $\sigma$ as \eqref{eq:sigmaConvention}.

\begin{figure}

\centering

\includegraphics[width = 8 cm]{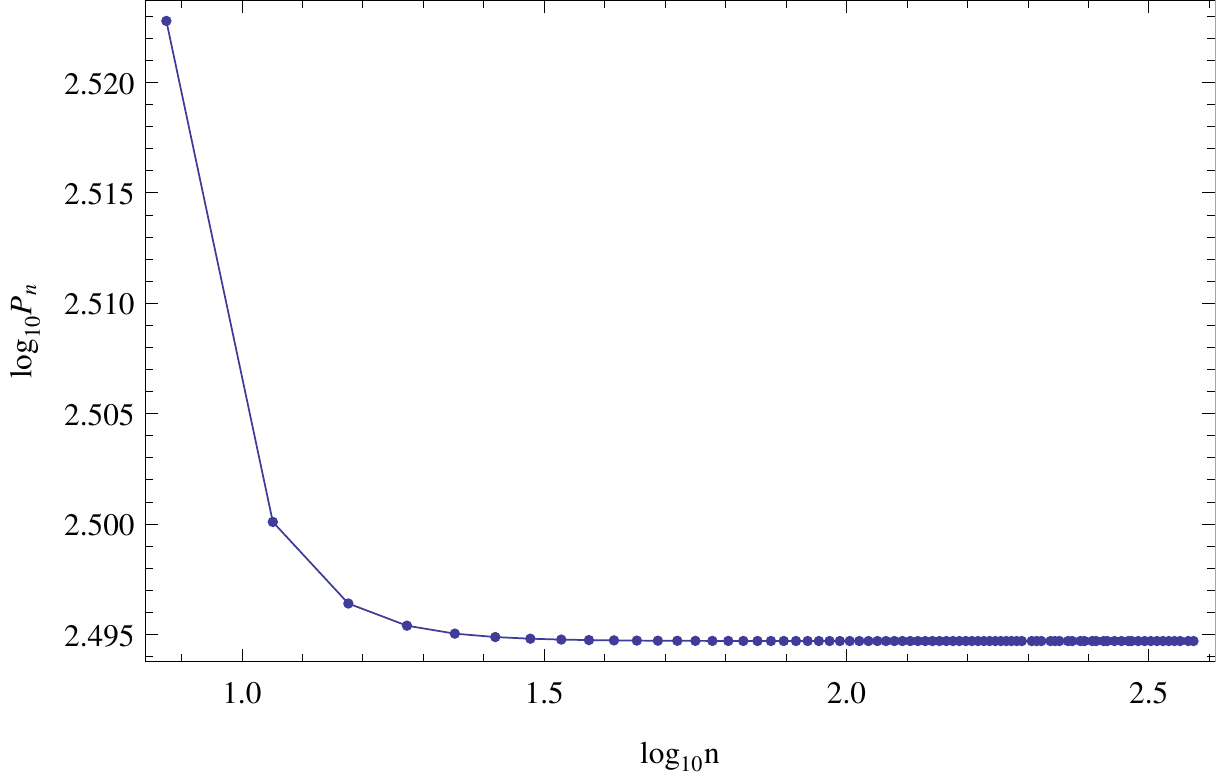}

\caption{The power spectrum obtained by numerically solving the perturbations with $\tilde{m} = 0$, $H_0 = \sqrt{8 \pi / 3}$.}

\label{fig:spectrum_0_1}

\end{figure}
\begin{figure}

\centering

\includegraphics[width = 8 cm]{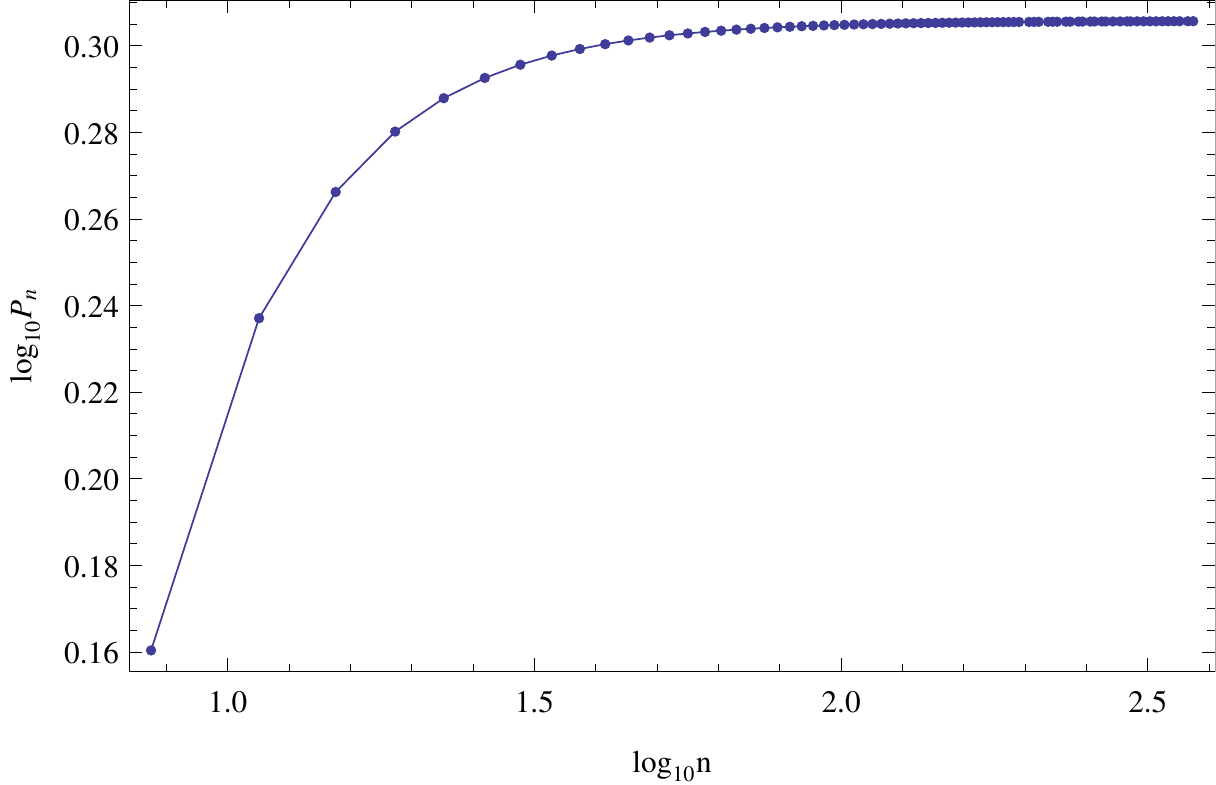}

\caption{The power spectrum obtained by numerically solving the perturbations with $\tilde{m} = 1000 \sqrt{0.1}$, $H_0 = \sqrt{8 \pi / 3}$.}

\label{fig:spectrum_1000_1}

\end{figure}
\begin{figure}

\centering

\includegraphics[width = 8 cm]{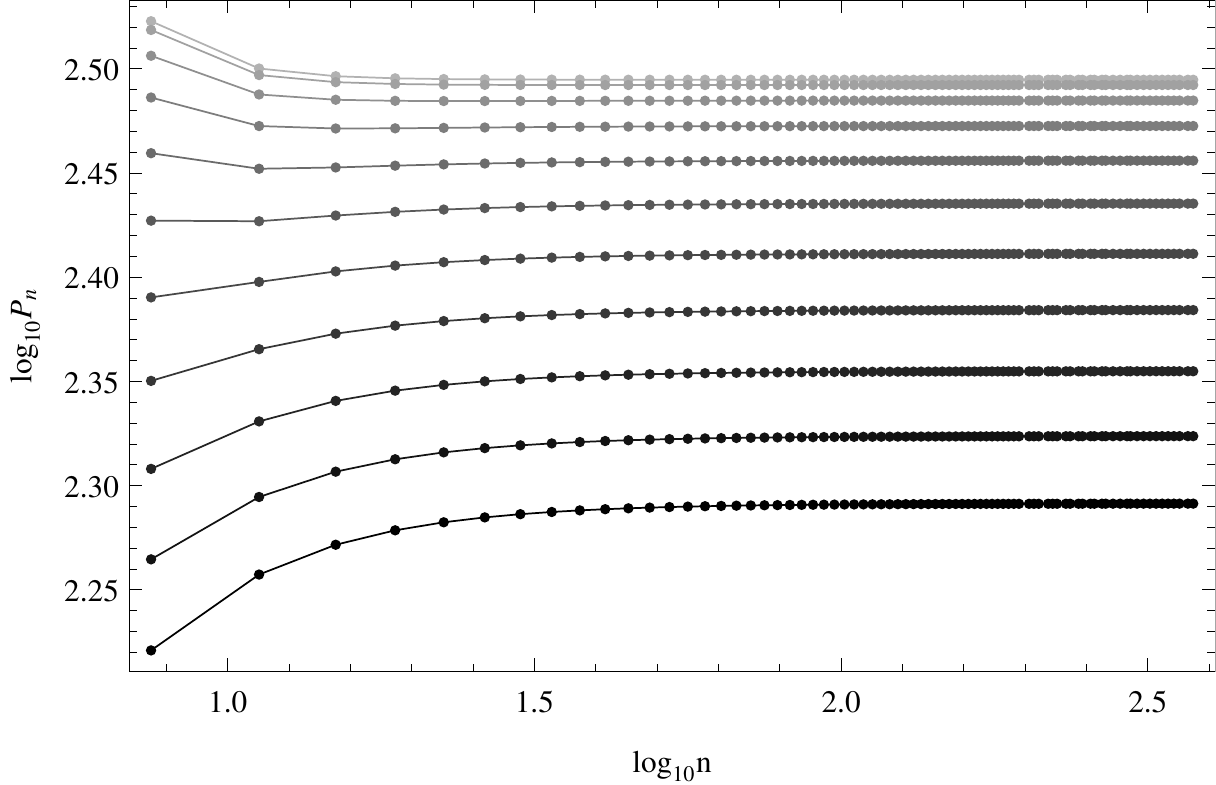}

\caption{The power spectrum obtained by numerically solving the perturbations with $\tilde{m} = \sqrt{0.1} \times \{ 0, 1, \dots, 10 \}$, from top to bottom. All spectra are plotted with $H_0 = \sqrt{8 \pi / 3}$.}

\label{fig:spectrum_m_1}

\end{figure}

	FIG.~\ref{fig:spectrum_0_1} shows the power spectrum in the massless case with $H_0 = \sqrt{8 \pi / 3}$. We see that while the power spectrum is scale-invariant in the small scales, it is enhanced in the large scales. FIG.~\ref{fig:spectrum_1000_1} is the power spectrum for a large mass $\tilde{m} = 1000 \sqrt{0.1}$ with $H_0 = \sqrt{8 \pi / 3}$. Opposed to the massless case, we see that in this massive case the large-scale spectrum is suppressed. In FIG.~\ref{fig:spectrum_m_1} we show the spectra corresponding to a range of masses, holding $H_0 = \sqrt{8 \pi / 3}$. We can observe the trend that, as the mass increases, the large-scale spectrum turns from being enhanced to being suppressed. We find that roughly the power is enhanced when $\tilde{m}$ is greater than $0.5 H_0$, and suppressed when $\tilde{m}$ is less than $0.5 H_0$.


\begin{figure}

\centering

\includegraphics[width = 8 cm]{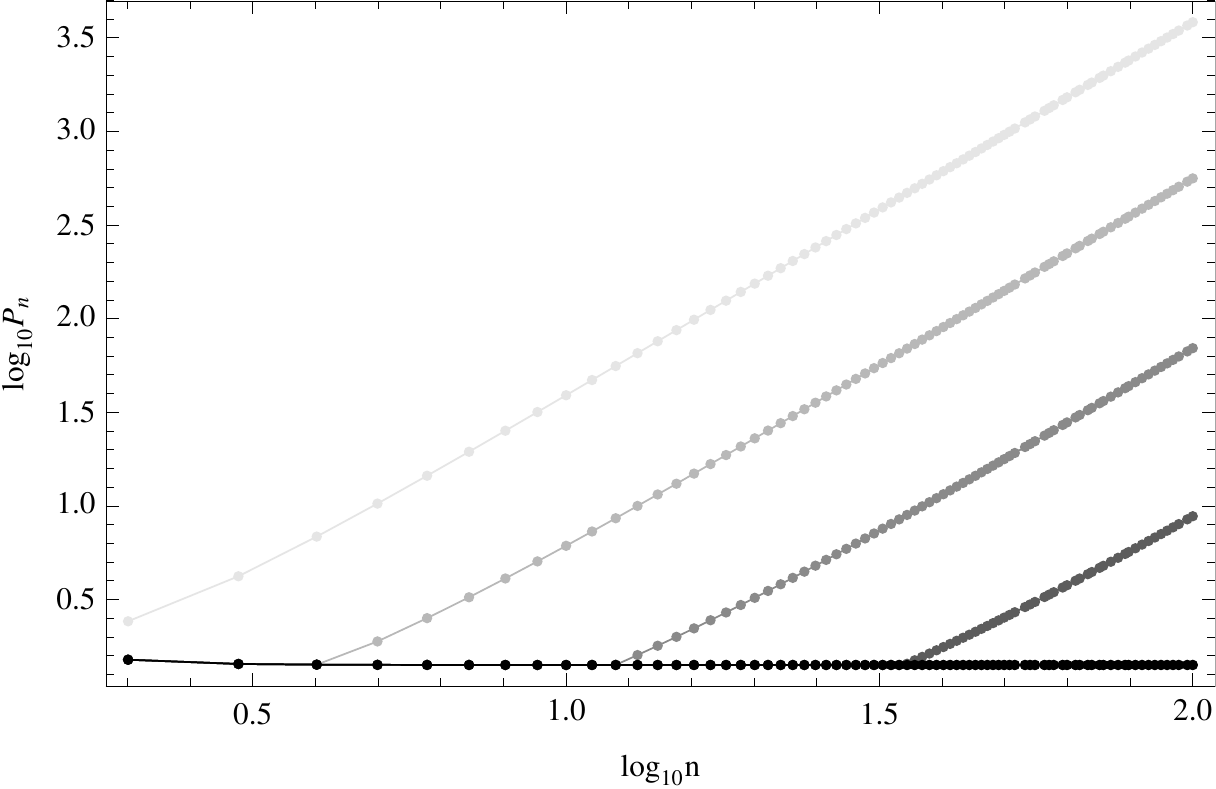}

\caption{The time evolution of power spectrum in the case of $\tilde{m} = 0$, $H_0 = \sqrt{8 \pi / 3}$. The darker curves correspond to the spectra at later times. The lightest curve is the initial Lorentzian spectrum at time $t_i$. For each $n$ mode, the power is evaluated up to its horizon crossing time.}

\label{fig:power_0_exit}

\end{figure}

\begin{figure}

\centering

\includegraphics[width = 8 cm]{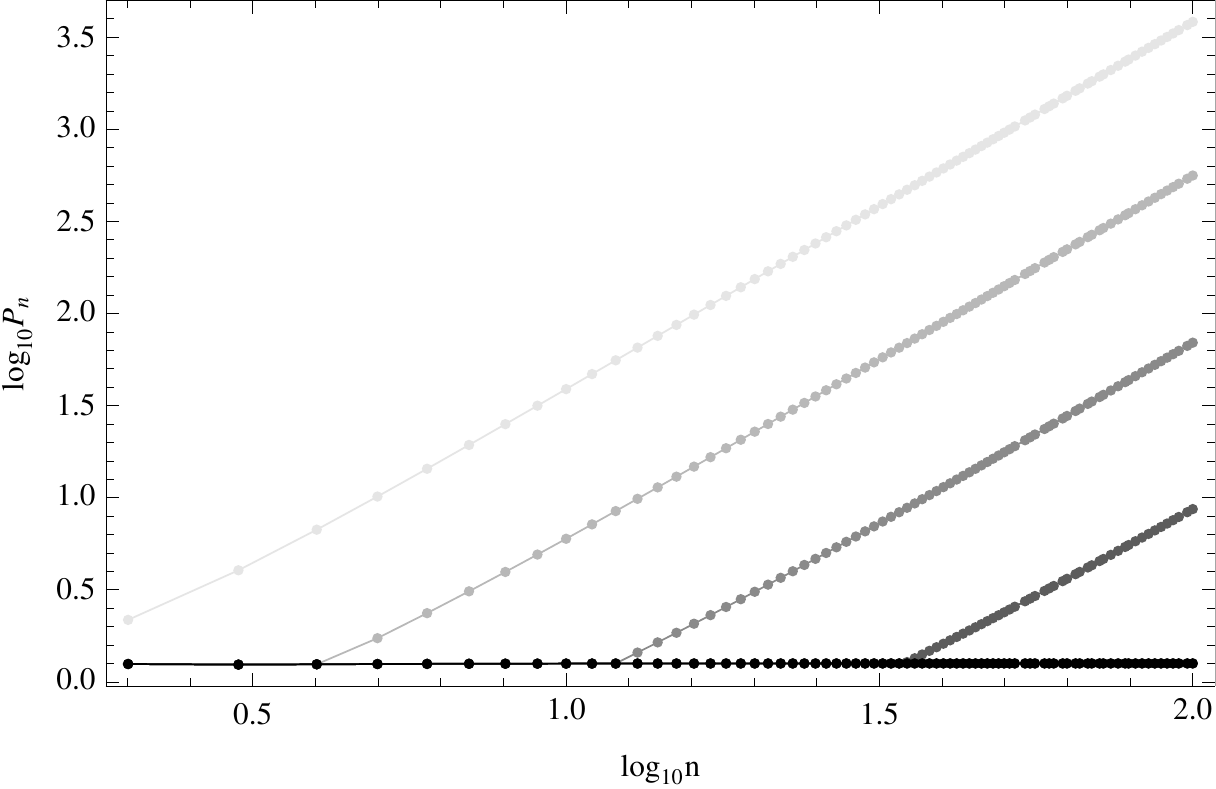}

\caption{The time evolution of power spectrum in the case of $\tilde{m} = 0.5 H_0$, $H_0 = \sqrt{8 \pi / 3}$. The darker curves correspond to the spectra at later times. The lightest curve is the initial Lorentzian spectrum at time $t_i$. For each $n$ mode, the power is evaluated up to its horizon crossing time.}

\label{fig:power_0p5H_exit}

\end{figure}

\begin{figure}

\centering

\includegraphics[width = 8 cm]{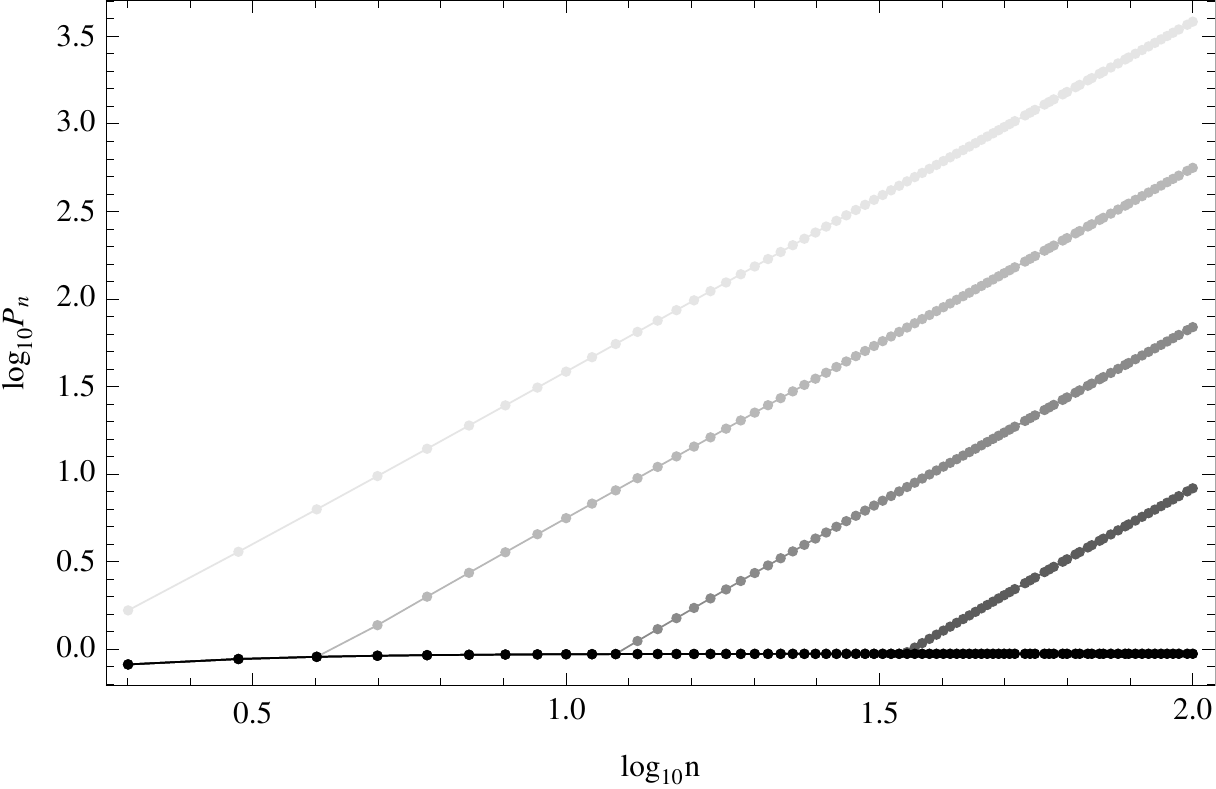}

\caption{The time evolution of power spectrum in the case of $\tilde{m} = H_0$, $H_0 = \sqrt{8 \pi / 3}$. The darker curves correspond to the spectra at later times. The lightest curve is the initial Lorentzian spectrum at time $t_i$. For each $n$ mode, the power is evaluated up to its horizon crossing time.}

\label{fig:power_H_exit}

\end{figure}

\begin{figure}

\centering

\includegraphics[width = 8 cm]{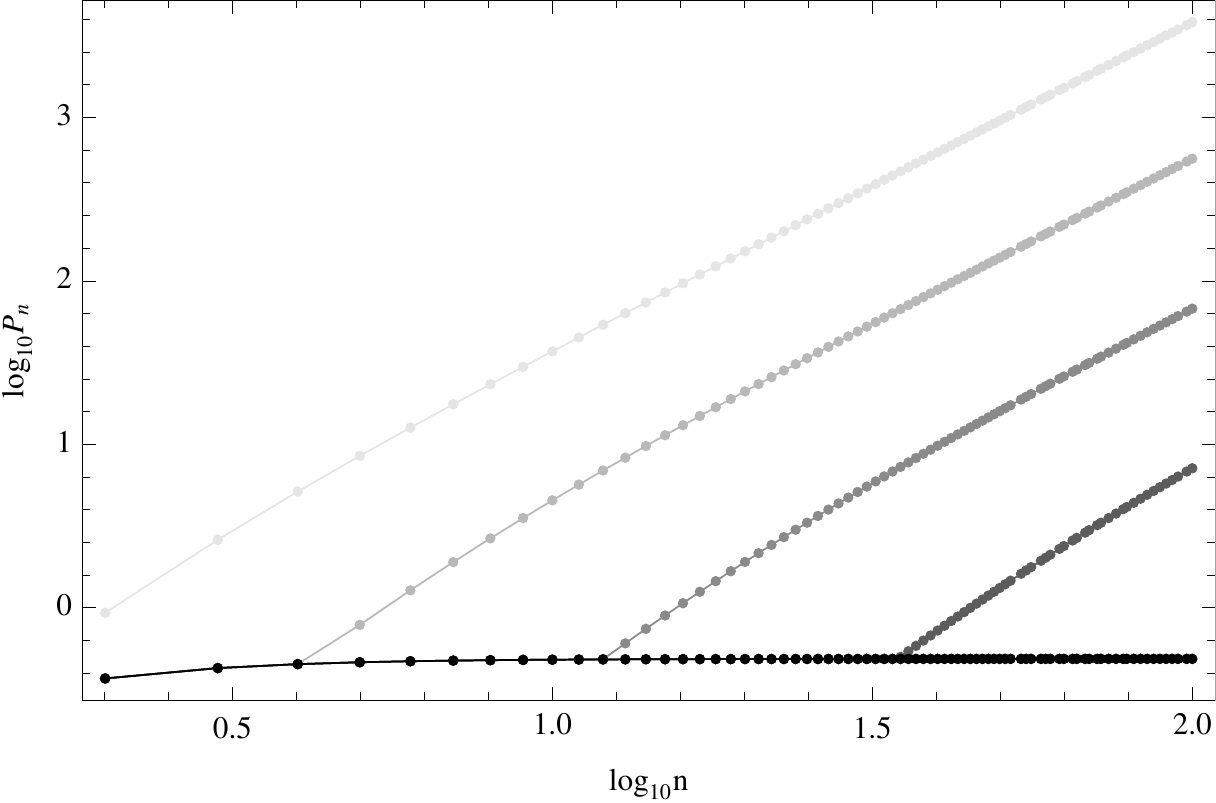}

\caption{The time evolution of power spectrum in the case of $\tilde{m} = 2 H_0$, $H_0 = \sqrt{8 \pi / 3}$. The darker curves correspond to the spectra at later times. The lightest curve is the initial Lorentzian spectrum at time $t_i$. For each $n$ mode, the power is evaluated up to its horizon crossing time.}

\label{fig:power_2H_exit}

\end{figure}

	To find out the mechanism that leads to this transition from enhancement to suppression as the mass increases, we first study the time evolution of the power spectrum in the Lorentzian space. The time evolution of spectrum in the massless case is given in FIG.~\ref{fig:power_0_exit}. For massive case, the time evolution of the spectra for the cases of $\tilde{m} = 0.5 H_0$, $H_0$, and $2 H_0$ is given in FIGs.~\ref{fig:power_0_exit} to \ref{fig:power_2H_exit}. 


	Through the spectrum evolution, we find that the power enhancement or suppression are reflected in the initial spectra in the Lorentzian space. At the small scales, before the horizon exit the slopes of the spectra are close to that of the spectrum of the Bunch-Davis vacuum. At the horizon crossing, the small-scale spectra are nearly scale-invariant. At the large scales, we see that at the horizon crossing the spectra is enhanced or suppressed determined by the mass of the scalar field as we showed before. Moreover, we note that even before the horizon crossing, already in the initial spectra in the Lorentzian space there are corresponding power enhancement or suppression relative to the small-scale Bunch-Davis vacuum. The origin of the power enhancement or suppression therefore lies on the Lorentzian initial condition, or, equivalently, on the Euclidean final spectrum.
	
	To find out the effect of mass on the Euclidean final spectrum, we note that the Euclidean equation of motion \eqref{eq:fEOMNoMetricEu} can also be analytically solved, yielding the solution
		\begin{align}
			\hat{f}_n(\tau) = A \frac{P_{\nu}^n[ \cos (H_0 \tau) ]}{\sin (H_0 \tau)},
		\end{align}
		where $A$ is an overall coefficient that has no effect on the final Euclidean spectrum,
		\begin{align}
			\nu = \frac{-1 + \sqrt{9 - 4 \tilde{m}^2 / H_0^2}}{2},
		\end{align}
		and we have picked the solution that is consistent with the no-boundary initial condition. When $\tilde{m}^2 / H_0^2 > 9 / 4$, $\nu$ and $\hat{f}_n(\tau)$ become complex. The power spectrum at the beginning of the Lorentzian time can be evaluated using the Euclidean solution at $\tau = \pi / 2 H_0$ through the boundary conditions. When evaluating the ratio $\tilde{f}_n / \dot{\tilde{f}}_n$ with complex $\tilde{f}_n$, we interpret it as the amplitude $| \tilde{f}_n / \dot{\tilde{f}}_n |$. We then have the initial power spectrum in the Lorentzian space as
		\begin{align}
			P(n) = \frac{3 n^3 H_0^2}{8 \pi} \left| \frac{P_{\nu}^n(0)}{P_{\nu}^n{}'(0)} \right|.
		\end{align}
		In the large-mass limit, we can intuitively understand the power suppression of the initial power spectrum induced by the mass term in the following way. In such a limit, the solution to the equation of motion \eqref{eq:fEOMNoMetricEu} roughly consists of an exponentially growing mode, $\exp(\tilde{m} \tau)$, and an exponentially decaying mode, $\exp(-\tilde{m} \tau)$. Hence, the amplitude $| \tilde{f}_n / \dot{\tilde{f}}_n |$ is roughly of the order of $1 / \tilde{m}$, which is suppressed by $\tilde{m} = m / \sqrt{V_0}$. Note that the large-mass limit actually lies beyond the linear regime of perturbations, and the purpose of considering it is only to provide an intuitive understanding. As shown in FIG.~\ref{fig:spectrum_m_1}, the long-wavelength spectrum is already suppressed as $\tilde{m}^2 / H_0^2$ is as small as roughly $0.1 \sqrt{6} / \sqrt{8 \pi / 3} \approx 0.43$. Therefore, it only requires a moderate mass to induce the effect of suppression.


\section{Conclusions and discussion}
\label{sec:Conclusion}

	In this paper, we investigated the power spectrum of perturbations due to the no-boundary wave function \cite{Hartle1983}. We have relied on very conservative approaches, such as the canonical quantization \cite{DeWitt1967}, Euclidean path integral approach and the steepest descent approximation \cite{Hartle1983}, use of instantons at the background as well as perturbation levels \cite{Halliwell1985}, and so on, which are consistent with traditional techniques of quantum field theory in several regimes \cite{Hwang2013a}.
	
	 What we can conclude are as follows. First, the inflationary universe is approximately scale-invariant for short-wavelength scales, while the power spectrum of the pure de Sitter space is enhanced for the long-wavelength scales. Therefore, our observation is definitely consistent with the scale-invariance of the Bunch-Davies vacuum for small scales, while the only difference is about long-wavelength modes as expected by the methods of quantum field theory \cite{Starobinsky1996}. Second, the power spectrum can be either enhanced or suppressed due to the detailed choice of the potential; for example, the mass term of the inflaton field. One can easily build a model including the mass term because its origin is nothing but the mass of the inflaton field. Our approximation still holds since the mass term maintains linear equations of motion as already discussed by Halliwell and Hawking \cite{Halliwell1985}. This opens a possibility that the power suppression is indeed a hint to that our universe starts from an instanton with a massive inflaton field that approximates the Hartle-Hawking wave function.
	
	There has been several alternative explanations about the CMB power suppression \cite{Chen2016}, but these explanations (e.g., phantomness or kinetic energy dominated era) have their own problems. On the other hand, in our approach, it is naturally consistent with the canonical quantization program without any \emph{ad hoc} assumption about the quantum state or matter contents. In this sense, our explanations are superior and conservative than the other approaches. It is also worthwhile to mention that, although it is not possible to claim that the power suppression confirms the Hartle-Hawking wave function, this work opens a possibility to confirm or falsify a theory of quantum gravity by investigating its effects through the experiments and observations. It also shows that the Euclidean quantum cosmology can expect observational contents with high precisions, against usual expectations (e.g., see \cite{Ashtekar2017}).
	
	This line of exploration definitely needs more work. It will be interesting to see more detailed calculations for realistic inflationary scenarios. For example, we investigated the quadratic potential for the inflaton \cite{Linde1983} only, but it can be easily extended to the Starobinsky-type inflation models \cite{Starobinsky1980}. Also, we investigated for compact and homogeneous instantons, but there are other instantons that also explain the origin of our universe; e.g., the Coleman-De Luccia instantons \cite{Coleman1980} or the Euclidean wormholes \cite{Chen2016b, Kang2017, Chen2017, Tumurtushaa:2018agq}. One more brave question is this: what is the relation between the big bounce model of the loop quantum cosmology \cite{Ashtekar2015} and the Hartle-Hawking wave function \cite{Hartle1983}? Both approaches explain the power suppression, but it is yet unclear which one is more suitable as the model of the beginning of our universe. We leave these interesting issues for future research topics.


\begin{acknowledgements}

	We are grateful for the discussions with Frederico Arroja, Jinn-Ouk Gong, and Antonino Marciano. P.~C.~and Y.~L.~are supported by Taiwan National Science Council under Project No.~NSC 97-2112-M-002-026-MY3 and by Taiwan National Center for Theoretical Sciences (NCTS). P.~C.~is in addition supported by U.S.~Department of Energy under Contract No.~DE-AC03-76SF00515. D.~Y.~is supported by the Leung Center for Cosmology and Particle Astrophysics (LeCosPA) of National Taiwan University (103R4000). D.~Y. is supported in part by the Korean Ministry of Education, Science and Technology, Gyeongsangbuk-do and Pohang City for Independent Junior Research Groups at the Asia Pacific Center for Theoretical Physics and the National Research Foundation of Korea (Grant No.: 2018R1D1A1B07049126).

\end{acknowledgements}


\end{document}